# Critical current density and microstructure of iron sheathed multifilamentary $Sr_{1-x}K_xFe_2As_2$/Ag composite conductors

Chao Yao[1], He Lin[1], Qianjun Zhang[1], Xianping Zhang[1], Dongliang Wang[1], Chiheng Dong[1], Yanwei Ma[1, a)], Satoshi Awaji[2] and Kazuo Watanabe[2]

[1] Key Laboratory of Applied Superconductivity, Institute of Electrical Engineering, Chinese Academy of Sciences, PO Box 2703, Beijing 100190, China

[2] High Field Laboratory for Superconducting Materials, Institute for Materials Research, Tohoku University, Sendai 980-8577, Japan

*E-mail: ywma@mail.iee.ac.cn

**Abstract**

Iron-based superconductors have been considered to be very promising in high-field applications, for which multifilamentary wire and tape conductors with high mechanical strength are essential. In this work, 7-， 19- and 114-filament $Sr_{0.6}K_{0.4}Fe_2As_2$ (Sr-122) superconducting wires and tapes with silver as matrix and iron as outer reinforcing sheath were produced by the *ex situ* powder-in-tube method. The mass densities of Sr-122 phase in 7- and 19-filament conductors were investigated by microhardness characterization, which revealed a positive correlation

---

[a)] Author to whom any correspondence should be addressed.

between hardness and transport critical current density ($J_c$) in round wires and flat tapes with various thicknesses. For Sr-122/Ag/Fe 114-filament conductors, in which an average cross-sectional filament size smaller than 50 microns was achieved by drawing into round wires of 2.0 mm in diameter, the transport $J_c$ can be significantly enhanced by flat rolling, as for the 7- and 19-filament conductors. The highest transport $J_c$ of the 7-, 19- and 114-filament Sr-122/Ag/Fe tapes reached $1.4 \times 10^4$, $8.4 \times 10^3$ and $6.3 \times 10^3$ A cm$^{-2}$ (4.2 K, 10 T), respectively, showing a $J_c$ degradation with the increase of filament number. This $J_c$ degradation can be ascribed to the sausage effect for filaments in longitudinal direction and the grain refinement in these very fine filaments.

## 1. Introduction

Since the discovery of iron-based superconductors [1], great efforts have been made to push them towards larger-scale applications [2, 3]. Iron pnictides AeFe$_2$As$_2$ (Ae = alkali or alkali earth elements), which are called 122 type iron-based superconductors, were found to have superconducting transition temperature ($T_c$) up to 37~38 K by potassium doping at the Ae site [4, 5], ultrahigh upper critical field ($H_{c2}$) above 100 T [6, 7], very small $H_{c2}$ anisotropy ($\gamma = H_{c2}//ab / H_{c2}//c$) about 1.5-2 [8, 9], and large critical current density ($J_c$) over $10^6$ A cm$^{-2}$ in thin films [10, 11]. These attractive properties make 122 iron pnictide superconductors to be very promising candidates for high-field applications, which need high-performance and low-cost wire and tape conductors with large current carrying capacity to generate high magnetic fields, sufficient mechanical strength to withstand the electromagnetic and thermal stress during operation, and fine superconducting filaments in metal matrix to protect against flux jumps and thermal quenching [12].

Nowadays the powder-in-tube (PIT) method has been proved to be a convenient and scalable way to develop iron-based superconducting wires and tapes, and silver is so far the most suitable sheath material for 122 iron pnictides, since it is chemically stable enough to prevent the generation of nonsuperconducting phase at the interface between sheath and 122 phase [13]. Recently, by hot and cold uniaxial pressing process to achieve highly densified superconducting phase, the transport $J_c$ in silver sheathed mono- and 7-filament Sr-122 and Ba-122 tapes has been greatly improved [14, 15]. At present, the highest $J_c$ value, which was achieved in hot pressed Sr-122 monofilamentary tapes, has reached $1.2 \times 10^5$ A cm$^{-2}$ at 4.2 K and 10 T [16]. However, from the viewpoint of practical applications mentioned above, when used as sheath material, the high price and poor mechanical properties of silver will hold back the large-scale applications of 122 iron pnictides [17], as in the case of BiSrCaCuO/Ag conductors. On the other hand, the oxygen permeability of sheath material during sintering process, which reduces the material choice for the sheath of BiSrCaCuO conductors to silver or some silver rich alloys [18], is not required for iron pnictides. Therefore, using other cheap and stiff metal material as the outer sheath for 122-pnicide/Ag composite conductors can be a practical proposal to reduce the ratio of silver cost, provide sheath chemical stability, and enhance mechanical properties at the same time. Previously, monofilamentary PIT Sr(Ba)-122/Ag composites conductors with iron [19, 20], copper [21, 22] and stainless steel [23] outer sheath have been prepared, respectively. On the other hand, we have successfully developed 7-filament Sr-122/Ag/Fe wires and tapes based on PIT method, and transport $J_c$ up to $3.3\times10^3$ A cm$^{-2}$ was obtained at 4.2 K and 10 T [24]. However, for high-field applications, conductors may perform in AC region. Therefore, multifilamentary conductors with large filament number and very fine filaments are needed in order to reduce the AC loss induced by the eddy currents in metallic matrix and the interfilamentary coupling currents [18]. Therefore, in this work, 7-, 19- and 114-filament Sr-122/Ag/Fe composite conductors were prepared.

For monofilamentary iron pnictide conductors, the main factors that determine the $J_c$ property are mass density, residual microcracks, impurity phases, and grain texture in superconducting cores [13]. However, for multifilamentary conductors with composite metal sheaths, we must consider more for $J_c$ performance due to their complex architecture and very small filament size. For BiSrCaCuO/Ag multifilamentary wires and tapes, microhardness were widely used to indicate the variations of mass density for filaments in different locations on transverse cross section, and provided a connection between deformation methods/parameters and $J_c$ distribution [25-30]. On the other hand, structural uniformity of BiSrCaCuO filaments along the longitudinal direction is also an important issue that must be taken into consideration. Because of different mechanical behaviors during cold deformation process, irregular interfaces often formed between silver sheath and BiSrCaCuO phase, causing sausage like filaments [27, 30-35]. These nonuniform filaments will lead to bottleneck effect on transport currents, and decrease the transport $J_c$ of the whole multifilamentary conductors. In this work, by microhardness and micromorphology characterizations, we investigate the influence of mechanical deformation and filament number on the microstructure and transport properties of multifilamentary Sr-122/Ag/Fe composite conductors, and give some suggestions for the $J_c$ improvements in the future.

**2. Experimental details**

The starting materials for synthesizing $Sr_{0.6}K_{0.4}Fe_2As_2$ precursor are small Sr pieces, K bulks and fine Fe and As powders. An excess of 20% K was added to compensate its loss during the sintering. The powders were mixed and ground by ball milling in Ar atmosphere for about 12 h, then sealed and sintered in an Nb tube for 35 h at 900 °C. The precursor was added with 5 wt.% Sn to improve grain connectivity, and then the mixture was ground into fine powder in an agate mortar in Ar atmosphere. The ground powder was packed into silver tubes (OD: 8 mm and ID: 5 mm), which were drawn into monofilamentary wires of 1.58, 1.21 and 0.95 mm in diameter. 7(19) short 1.58(0.95)-mm-diameter monofilamentary Sr-122/Ag wires were then bundled into iron tubes (OD: 8 mm and ID: 5 mm), and finally deformed into 7(19)-filament Sr-122/Ag/Fe wires of 2 mm in diameter and tapes with various thicknesses by drawing and flat rolling. For 114-filament Sr-122/Ag /Fe wires, 19 short 1.21-mm-diameter monofilamentary Sr-122/Ag wires were bundled into a silver tube (OD: 8 mm and ID: 6.4 mm), which was drawn into a 19-filament wire of 1.58 mm in diameter. 6 short 19-filament Sr-122/Ag wires and a short silver wire (also with 1.58 mm in diameter) in the center were bundled into an iron tube (OD: 8 mm and ID: 5 mm), and finally deformed into 114-filament Sr-122/Ag/Fe wires of 2 mm in diameter and tapes with various thicknesses by drawing and flat rolling. After cold deformation process, all the multifilamentary conductor samples were submitted to a heat treatment at 900 $^{\circ}$C for about 0.5 hours.

Transport critical currents were measured at 4.2 K using short wire and tape samples of 3 cm in length by the standard four-probe method with an evaluation criterion of 1 $\mu$V cm$^{-1}$. The applied fields in transport critical current $I_c$ measurement are parallel to the tape surface. Transport $J_c$ was obtained by dividing current values through the cross section area of superconducting cores. Resistance measurements were carried out on a Quantum Design's physical property measurement system (PPMS) using the four-probe method. The sample resistivity is defined as $\rho=RS/L$ ($R$ is the value of measured resistance, $S$ is the area of transverse cross section for the whole sample including Sr-122, silver matrix and iron sheath, and $L$ is the distance between the two voltage leads). The Vickers hardness of sheath materials and Sr-122 filaments was measured on the polished transverse cross section of samples with 0.05 kg load and 10 s duration after heat treatment. The micromorphology for Sr-122 phase was examined by a Zeiss SIGMA scanning electron microscope (SEM) after peeling off the iron sheath and part of silver matrix. The element composition of Sr-122 phase was determined by an energy dispersive x-ray spectroscopy (EDX) on SEM.

## 3. Results and discussion

Figure 1 shows the transverse cross sections for 7- and 19-filament Sr-122/Ag/Fe wires of 2.0 mm in diameter and tapes of 0.6 mm in thickness, as well as the average values of Vickers hardness for Sr-122 filaments indicated by different colours. Being much more convenient than the dierect measurement of mass density for superconducting cores, hardness can be used to monitor the evolution of local mass density during the manufacturing process [28]. The 7- and 19-filament Sr-122/Ag/Fe round wire shows relatively uniform core densities with Vickers hardness of 49~63 and 38~69, respectively, as shown in figure 1(a) and (b). In figure 1(c) and (d), due to the defomation induced densification, both 7- and 19-filament tapes flat rolled to 0.6 mm in thickness show much higher hardness compared with round wires. The mass redistributions of Sr-122 phase after rolling for 7- and 19-filament tapes are similar: the lowest value of hardness (88 for 7-filament tape and 84 for 19-filament tape) was measured in the filaments in the center area of the tapes, and the filaments near the silver sheath have higher hardness up to 107 and 133 for 7- and 19-filament tapes, respectivly. This result is similar to the flat rolled 7-filament BiSrCaCuO/Ag tapes, in which the filament closer to the sheath show higher average hardness, though the microhardness profiles across each individual filament are different [26]. In addition, we also measured the hardness for silver matrix and iron sheath for all the heat treated Sr-122/Ag/Fe conductors, and the results show no big difference for all the samples. As listed in table 1, the hardness for silver matrix is 30~39, similar to the values in sintered Bi-2223/Ag tapes [36], and the hardness for iron sheath is 91~97, higher than silver matrix. Therefore, the mechanical strength for the whole conductors can be significantly improved by iron sheaths.

Figure 2 presents the evolution of hardness for each Sr-122 filament in 7- and 19-filament conductors and the average transport $J_c$ (in self-field and 10 T) for the Sr-122 filaments when the conductors were rolled from round wires into flat tapes. At 4.2 K and 10 T, the highest transport $J_c$ reached $1.4\times10^4$ and $8.4\times10^3$ A cm$^{-2}$ for 7- and 19-filament Sr-122/Ag/Fe tapes, respectively. Compared with the 7-filament Sr-122/Ag/Fe conductors, the 19-filament conductors has a lower mass homogeneity for Sr-122 filaments, but in general their mass density and transport $J_c$ both increase with the reduction of tape thickness. This result is in accordance with the case of Ba-122/Ag and Ba-122/Ag/SS monofilamentary tapes, showing a positive correlation between hardness and transport $J_c$ [23].

The transverse cross sections for the 114-filament Sr-122/Ag/Fe wires and tapes are presented in figure 3. As shown in figure 3(a), filaments with size smaller than 50 μm were achieved in round wires with of 2.0 mm in diameter. The field dependence of the transport $J_c$ for 114-filament Sr-122/Ag/Fe wire of 2.0 mm in diameter and tapes with various thicknesses is plotted in figure 4. The transport $J_c$ of round wires is about 800 A cm$^{-2}$ (4.2 K, 10 T), and when they are flat rolled into tapes, the transport $J_c$ gradually grows with the reduction of tape thickness. Finally the $J_c$ value comes up to $6.3\times10^3$ A cm$^{-2}$ (4.2 K, 10 T) in 0.6 mm thick tapes. Though the filaments are too small for hardness measurement for the 114-filament conductors, the mechanical deformation induced densification for the Sr-122 phase can be inferred since the transport $J_c$ can be significantly enhanced by flat rolling as for the 7- and 19- filament conductors. It is noteworthy that these 114-filament samples shows quite weak field dependence up to 14 T, though they have different sizes and shapes, indicating very promising potential in high-field applications.

On the other hand, as compared in figure 5, it is clear that the transport $J_c$ decreases with the increse of filament numbers, i.e. the reduction of filament size in the whole field region for 7-, 19- and 114-filament Sr-122/Ag/Fe tapes with 0.6 mm in thickness. At 4.2 K and 10 T, the transport $J_c$ for 114-filament tapes is $6.3\times10^3$ A cm$^{-2}$, nearly half of the $1.4\times10^4$ A cm$^{-2}$ for 7-filament tapes. Therefore, detailed microstructural characterizations are needed to explain the $J_c$ degradation of 19- and 114-filament conductors.

Before any investigations on microstructure, in order to check the crystallinity of Sr-122 phase, the superconducting transition for 7-, 19- and 114-filament Sr-122/Ag/Fe wires and tapes was examined by $R$-$T$ measurements from 10 to 300 K, as presented in figure 6. The onset and zero superconducting transition temperature ($T_{c,\,onset}$ and $T_{c,\,zero}$) and resistivity at 300 K ($\rho_{300}$) for the samples are summarized in table 1. All the wire and tape samples show $T_{c,\,onset}$ above 33 and 34 K, respectively, and also very steep superconducting transitions with transition widths smaller than 1.5 K. Compared with the wire samples, the 0.6 mm thick tape samples with the same filament numbers show an increase about 1.0 K for both $T_{c,\,onset}$ and $T_{c,\,zero}$, indicating the deformation induced enhancement for the density of Sr-122 phase, since a highly dense superconducting phase is beneficial to its reaction and formation during the heat treatment, thus results in a high $T_c$. Due to the increased ratio of sheath materials to Sr-122 phase, the resistivity in the normal state decreases with the increase of filament number for both wires and tapes. For the wires and tapes with the same filament number, the latter ones exhibit lower resistivities in the normal state, which is the result of improved contact of silver and iron sheath materials [24]. Based on the analysis of superconducting transition above, in this work the Sr-122 phase exhibits similar superconducting transitions in wires/tapes with different filament numbers.

The optical micrographs of the longitudinal cross sections of polished 7-, 19- and 114-filament Sr-122/Ag/Fe tape are presented in figure 7. In figure 7(a), we can see that the filament size of 7-filament Sr-122/Ag/Fe tape is very uniform in the tape axis direction. In figure 7(b), for the 19-filament Sr-122/Ag/Fe tape, in contrast to the uniform filament B1, filament B2 shows a wavy interface between the filament and sheath. The narrowed Sr-122 region marked with dashed box on filament will restrict the pass of larger currents. In 114-filament tape shown in figure 7(c), though some uniform filaments like filament C1 can be observed, some others like filament C2 suffer from serious sausage effect due to the different hardness between the sheath and very fine filaments. As marked with arrows in filament C2, the necked regions will cause largely reduced or even discontinuous current path.

In order to characterize the micromorphology of the Sr-122 grains inside our multifilamentary tapes, 7-, 19- and 114-filament Sr-122/Ag/Fe tape of 0.6 mm in thickness were peeled off their iron sheath and part of silver matrix for SEM observation. In figure 8(d), the inset shows a low-magnification SEM image of 114-filament tape after removing the sheath. We can clearly see some residual Sr-122 filaments and silver matrix. The red rectangular box marks the area in which the high-magnification SEM image in the inset of figure 8(c) and the EDX spectrum in figure 8(d) were obtained. The EDX analysis comfirms that the area is consisted of Sr-122 phase. Like the inset of figure 8(c), the insets of figure 8(a) and (b) show the micromorphology of Sr-122 phase for 7- and 19-filament tapes, respectively. Compared with the 7- and 19-filament tapes which contain inhomogeneous Sr-122 grains and their agglomerates, the 114-filament tape has many smaller Sr-122 grains. The distributions of their size and the corresponding Gauss fit curves for 7-, 19- and 114- filament tapes are plotted in figure 8(a), (b) and (c), and the peaks of the fitting curves are at 2.0, 1.6 and 1.3 μm, respectively. This result indicates that grain refinement occurs due to the crushing effect in deformation process with the reduction of filament size. For 114-filament tapes, Sr-122 particles larger than 4 μm were significantly eliminated. The refined grains can increase the density of grain boundaries and reduce the degree of grain texture, which are not beneficial to the $J_c$ improvement, as for high-temperature cuprate superconductors [12]. Therefore, the $J_c$ degradation in 19- and 114-filament Sr-122/Ag/Fe conductors can be ascirbed to the nonuniform filaments in longitudinal direction and the grain refinement effect.

According to the results presented above, there are still lagre room for the $J_c$ improvement of multifilamentary 122 type iron pnictide conductors through three factors. Fisrt, in monofilamentary Sr(Ba)-122/Ag tapes, it has been proved that the transport $J_c$ can be enhanced with the reducing of rolling thickness down to 0.30~0.26 mm [15, 37]. A large transport $J_c$ of $5.4\times10^4$ A cm$^{-2}$ (4.2 K, 10 T) with a high average Vickers hardness of 136 in superconducting core has been reported in flat rolled thin

Ba-122/Ag tapes of 0.3 mm in thickness [38]. Employing hot or cold pressing process, hardness over 150 and high $J_c$ up to $1.2\times10^5$ A cm$^{-2}$ (4.2 K, 10 T) can be obtained [15, 16], but such processes are not suitable for producing long-length tapes. In figure 2, we can see that the average hardness for Sr-122 filaments in 0.6 mm thick 7- and 19-filament Sr-122/Ag/Fe tape is just around 100, so we can expect thin multifilamentary tapes with higher transport $J_c$. However, due to the different hardness and workability of sheath materials, the deformation process suitable for such mechanically reinforced composites conductors is still needed to develop. Second, the sausage effect, which is a commonly observed phenomenon in BiSrCaCuO multifilamentary thin wires and tapes, can be alleviated by optimizing the reduction rate of deformation, the diameter of rollers, particle size of precursor, hardness of sheath materials and so on [31, 32, 34, 35]. Third, the grain refinement in thin filaments can enhance grain boundary pinning but at the same time suppress intergrain currents. For MgB$_2$ superconductors, in which grain boundaries are transparent for current flow, the grain refinement has an overall positive effect to $J_c$ performance [39, 40]. However, for iron pnictide superconductors, whose transport $J_c$ can be suppressed by high-angle grain boundaries, we may try to further improve the grain connectivity by optimizing heat treatment. Therefore, the configuration and fabricating process for multifilamentary Sr-122/Ag/Fe composite conductors is still needed to improve. If high hardness, uniform interface and well-connected grains can be achieved for filaments, high transport $J_c$ in multifilamentary iron-based superconductors can be achieved in the future.

## 4. Conclusion

In summary, 7-, 19- and 114-filament Sr-122/Ag/Fe superconducting wires and tapes were fabricated based on PIT method with transport $J_c$ up to $1.4\times10^4$, $8.4\times10^3$ and $6.3\times10^3$ A cm$^{-2}$ (4.2 K, 10 T), respectively. By hardness characterization, a positive correlation between hardness and transport $J_c$ was revealed for 7- and 19-filament Sr-122/Ag/Fe tapes. On the other hand, the degraded uniformity of interface between Sr-122 filaments and silver matrix, in addition to the grain refinement in very fine filaments were found to be the reason for the degradation of transport $J_c$ in 19- and 114-filament Sr-122/Ag/Fe tapes. Further optimization for the process of multifilamentary Sr-122/Ag/Fe conductors is still needed to achieve filaments with high mass density and advantageous microstructure so as to improve transport $J_c$.

**Acknowledgments**

This work is partially supported by the National '973' Program (grant No. 2011CBA00105), the National Natural Science Foundation of China (grant Nos. 51172230, 51202243 and 51320105015), and the Beijing Municipal Science and Technology Commission (grant No. Z141100004214002).

**Captions**

Figure 1 Optical images of the transverse cross section for 7- and 19-filament Sr-122/Ag/Fe wires of 2.0 mm in diameter and tapes of 0.6 mm in thickness. The average Vickers hardness of filaments is indexed by colours.

Figure 2 Evolution of the hardness for Sr-122 filaments and their average transport $J_c$ (at 4.2 K, in self-field and 10 T) with flat rolling for 7- and 19-filament Sr-122/Ag/Fe conductors

Figure 3 Optical images of the transverse cross section for 114-filament Sr-122/Ag/Fe (a) wires of 2.0 mm in diameter and tapes of (b) 1.0 mm and (c) 0.6 mm in thickness, showing the rearrangement of the Sr-122 filament during flat rolling.

Figure 4 Field dependence of the transport $J_c$ for 114-filament Sr-122/Ag/Fe wire of 2.0 mm in diameter and tapes with various thicknesses.

Figure 5 Field dependence of the transport $J_c$ for 7-, 19- and 114-filament Sr-122/Ag/Fe tapes of 0.6 mm in thickness.

Figure 6 Temperature dependence of the resistivity for 7-, 19- and 114-filament Sr-122/Ag/Fe wires of 2.0 mm in diameter and tapes of 0.6 mm in thickness. The insets show the enlarged part near the superconducting transition.

Figure 7 Optical images of the polished longitudinal cross sections for (a) 7-, (b) 19- and (c) 114-filament Sr-122/Ag/Fe tape of 0.6 mm in thickness. The regions that restrict transport currents are marked with a dashed box and arrows.

Figure 8 Gauss fits of the particle size distribution (by measuring over 200 particles

from SEM images) for the Sr-122 filaments in (a) 7-, (b) 19- and (c) 114-filament Sr-122/Ag/Fe tape of 0.6 mm in thickness. The insets show the typical SEM micromorphology of the Sr-122 filaments for the three tape samples. The inset of (d) is a SEM image showing the internal structure of a 114-filament tape after peeling off the iron sheath and part of silver matrix. The red rectangular box marks the area in which the magnified SEM observation for the inset of (c) and the EDX spectrum in (d) were performed.

TABLE I. Summary of the sample information.

| Sample shape | Diameter / Thickness (mm) | Filament number | Core hardness (Hv) | Sheath hardness (Hv) | | $T_{c,\,onset}$ (K) | $T_{c,\,zero}$ (K) | $\rho_{300}$ ($\mu\Omega$cm) | Transport $J_c$ (A/cm$^2$) (at 4.2 K, 10 T) |
|---|---|---|---|---|---|---|---|---|---|
| | | | | Ag | Fe | | | | |
| Wire | 2.0 | 7 | 49~63 | 37 | 95 | 33.5 | 32.2 | 13.8 | - |
| | 2.0 | 19 | 38~69 | 32 | 92 | 33.9 | 32.6 | 9.8 | - |
| | 2.0 | 114 | - | 33 | 93 | 33.9 | 32.6 | 8.3 | $8.0\times10^2$ |
| Tape | 0.6 | 7 | 88~107 | 33 | 92 | 34.4 | 33.1 | 10.5 | $1.4\times10^4$ |
| | 0.6 | 19 | 84~133 | 30 | 91 | 34.9 | 33.4 | 8.1 | $8.4\times10^3$ |
| | 0.6 | 114 | - | 39 | 97 | 35.1 | 33.9 | 7.2 | $6.3\times10^3$ |

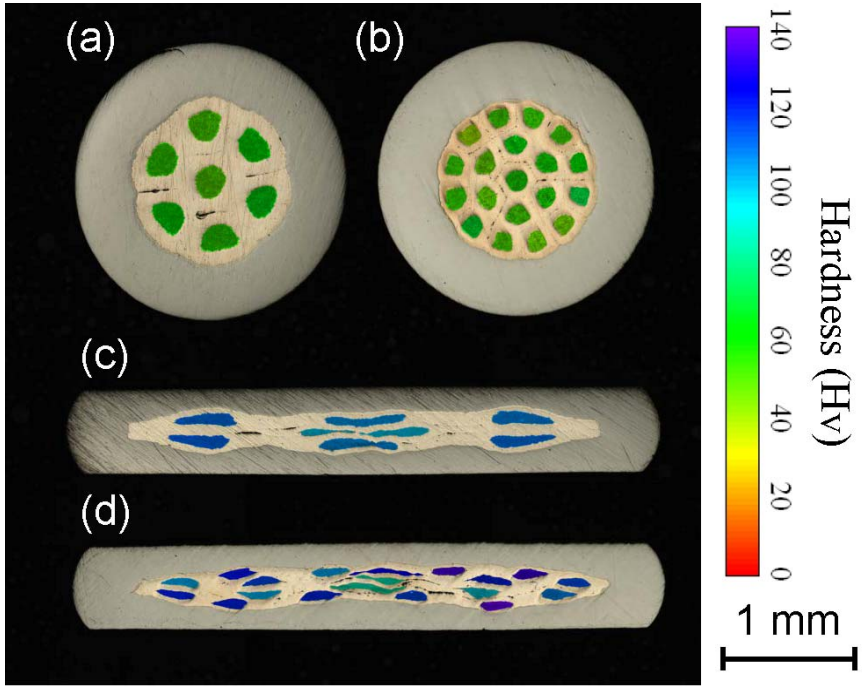

Figure 1 Yao et al.

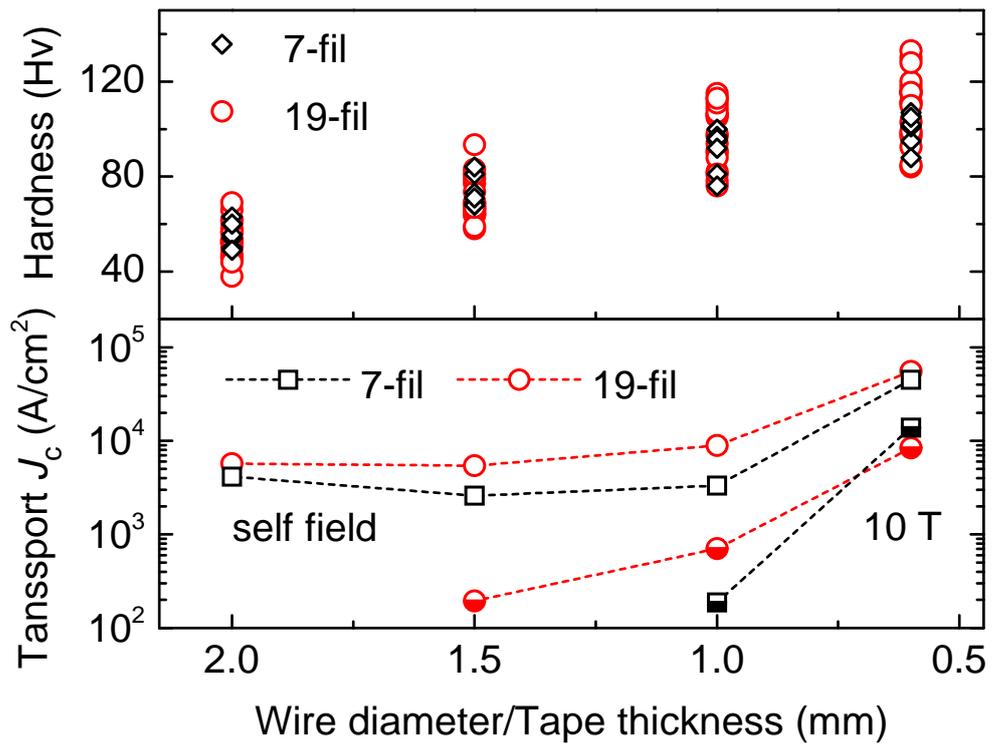

Figure 2 Yao et al.

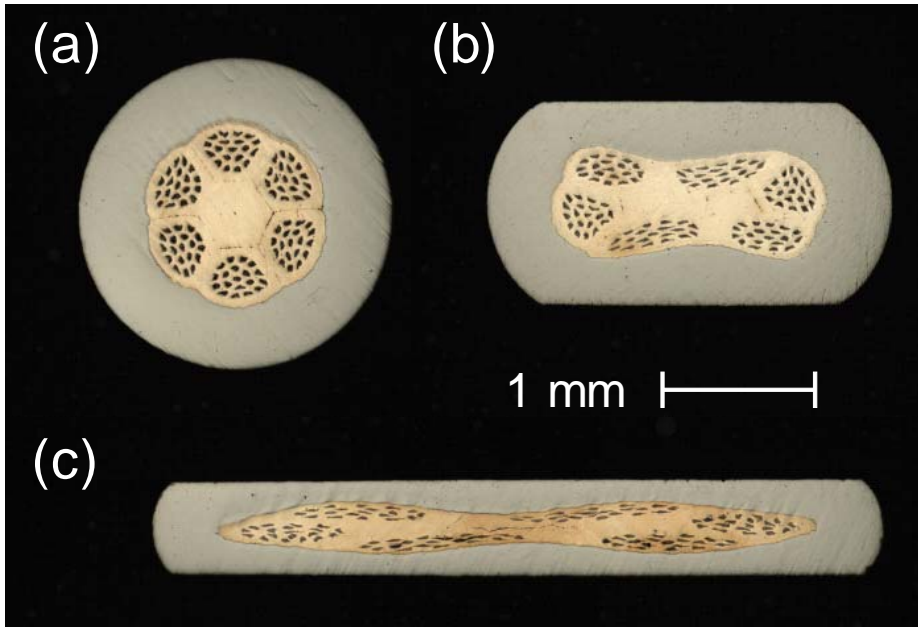

Figure 3 Yao et al.

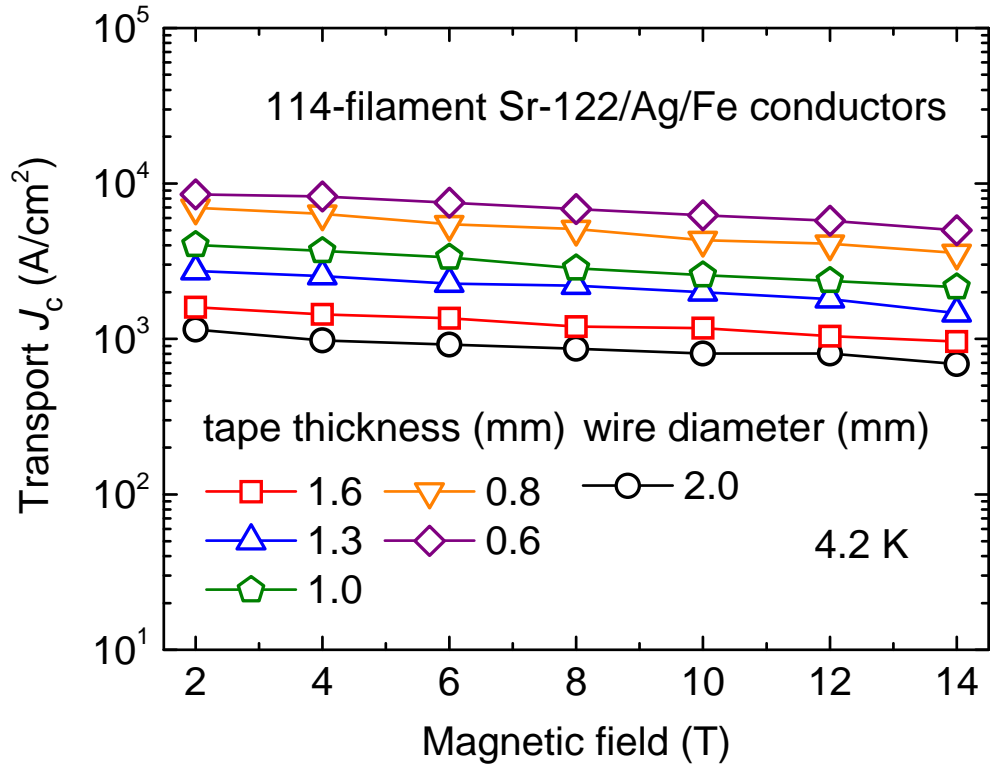

Figure 4 Yao et al.

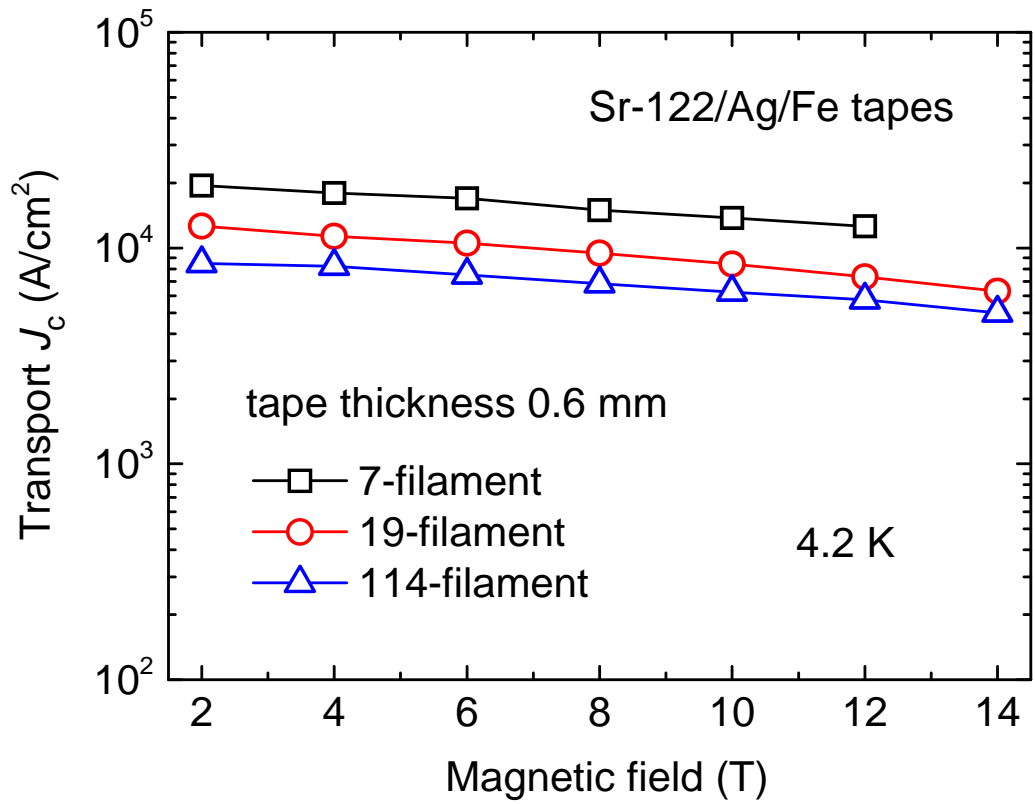

Figure 5 Yao et al.

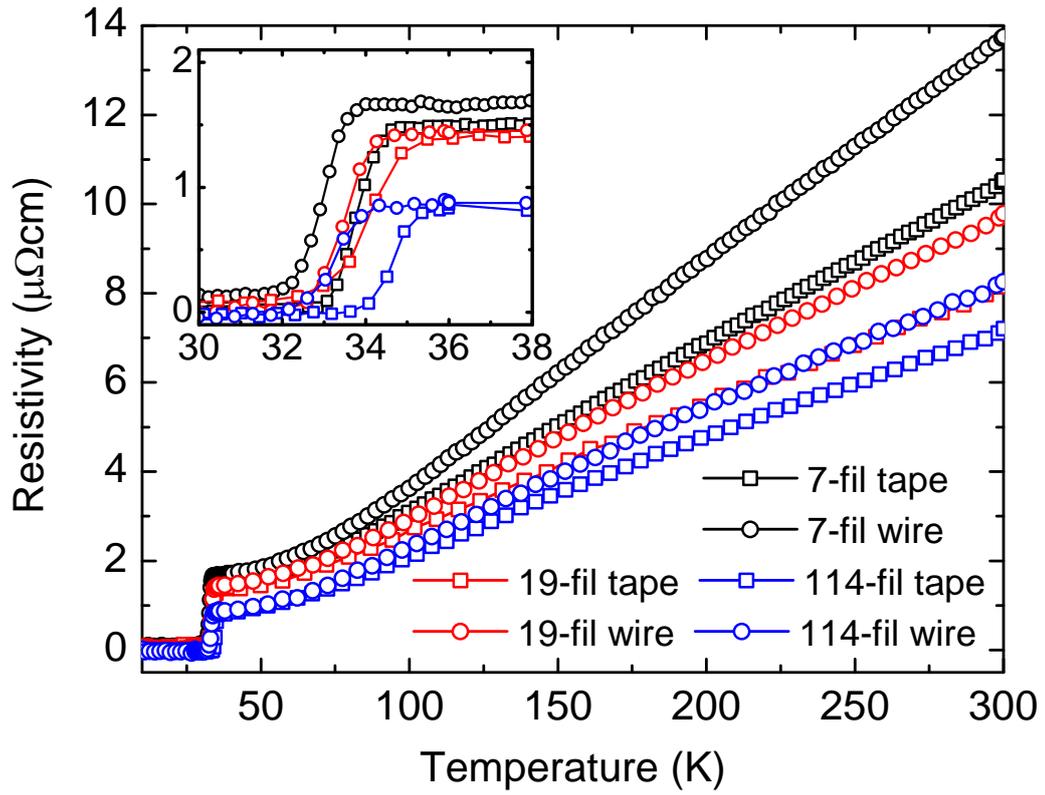

Figure 6 Yao et al.

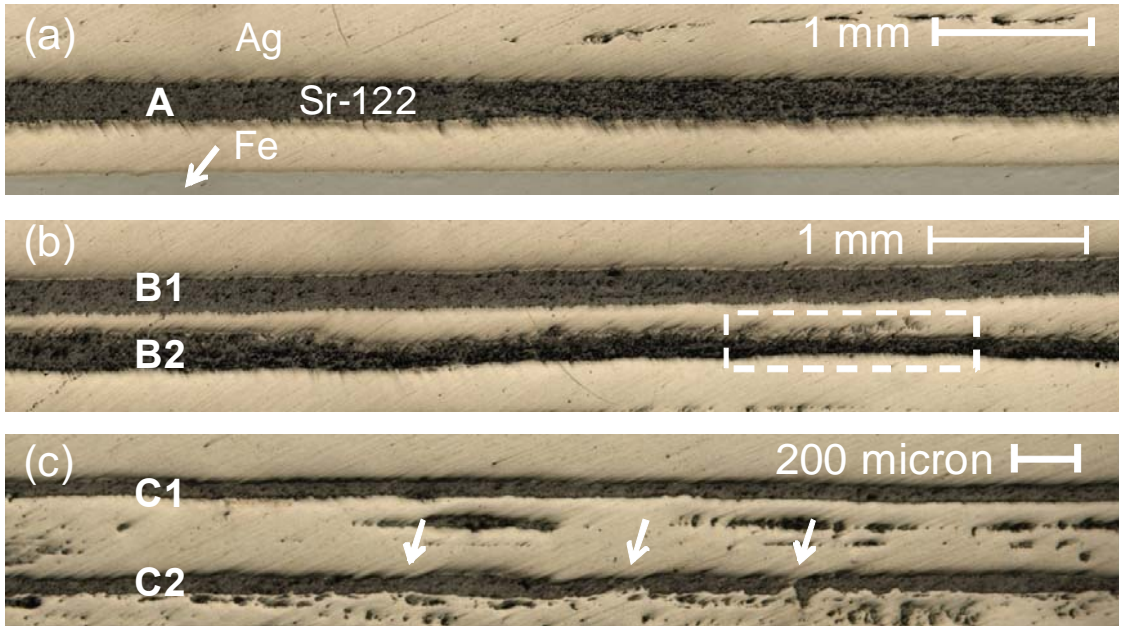

Figure 7 Yao et al.

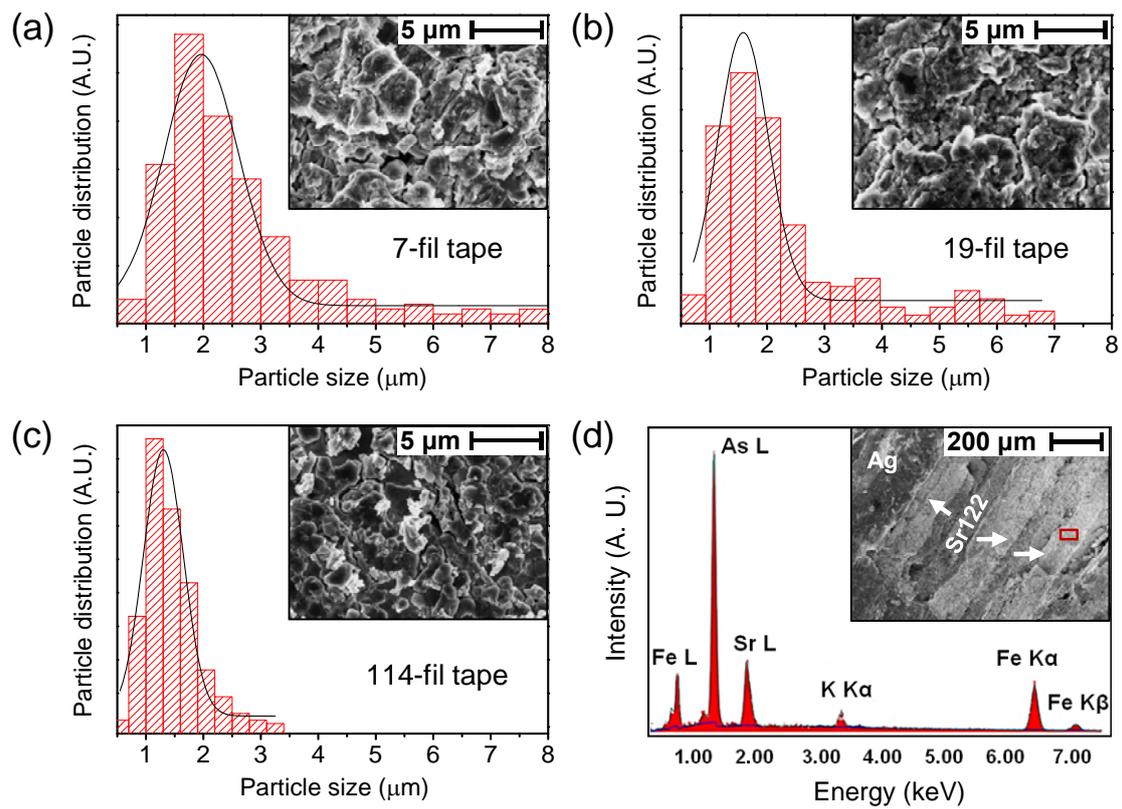

Figure 8 Yao et al.